# Identification of long-term concept-symbols among citations: Can documents be clustered in terms of common intellectual histories?

**Jordan A. Comins**[1,*] and **Loet Leydesdorff**[2]

**Abstract**

"Citation classics" are not only highly cited, but also cited during several decades. We test whether the peaks in the spectrograms generated by Reference Publication Years Spectroscopy (RPYS) indicate such long-term impact by comparing across RPYS for subsequent time intervals. Multi-RPYS enables us to distinguish between short-term citation peaks at the research front that decay within ten years versus historically constitutive (long-term) citations that function as concept symbols (Small, 1978). Using these constitutive citations, one is able to cluster document sets (e.g., journals) in terms of intellectually shared histories. We test this premise by clustering 40 journals in the Web of Science Category of Information and Library Science using multi-RPYS. It follows that RPYS can not only be used for retrieving roots of sets under study (cited), but also for algorithmic historiography of the citing sets. Significant references are historically rooted symbols among other citations that function as currency.

---

\* [1]*corresponding author*; Center for Applied Information Science, Virginia Tech Applied Research Corporation, Arlington, VA, United States; jcomins@gmail.com
[2] Amsterdam School of Communication Research (ASCoR), University of Amsterdam
PO Box 15793, 1001 NG Amsterdam, The Netherlands; loet@leydesdorff.net

## 1. Introduction

Current research builds upon prior discourse within a scientific community. At present, a primary means of demonstrating the applicability of previous work is through citations. Citations can be considered as functional linkages between ongoing scientific efforts with prior endeavors (de Solla Price, 1965; Garfield, Malin, & Small, 1978; Radicchi, Fortunato, & Castellano, 2008). The accumulation of citations is therefore commonly used as a proxy for quantifying the scholarly impact of research (Kostoff & Shlesinger, 2005; Marx, Bornmann, Barth, & Leydesdorff, 2014; Van Raan, 2000).

As science progresses, the value of early seminal works often becomes obscured, caused by the under-citation of historically important papers. This can be attributed to two processes: (1) important early works being incorporated into other documents such as textbooks, also known as obliteration-by-incorporation (Cozzens, 1985; E Garfield, 1975; Price, 1970) and (2) misattribution of a reference (Marx & Bornmann, 2013). Marx *et al.* (2014) introduced a novel technique for identifying historically important papers called Reference Publication Year Spectroscopy (RPYS). RPYS reconstructs the history of a field and allows for the detection of early seminal works by mining the aggregated reference sections associated with a carrying set of publications. Specifically, by capturing the publication years of the references cited, RPYS creates a frequency distribution of reference years cited by a user-chosen set of publications (Marx et al., 2014). Using a procedure to normalize the data over time, this distribution can be altered so that one controls for the empirical observation that articles tend to cite more recent literature (Marx, 2011). The normalization procedure thus controls for temporal-biases while the selection of the set of publications to mine in the first-place keeps the analysis confined to a specific user-generated topic area or community.

RPYS yields a single vector representing the normalized number of references cited by a set of citing publications over time. When visualized as a time-series, this vector tends to emphasize years where relatively significant findings were published with the appearance of peaks (for example: graphene and solar cells, Marx et al., 2014; the misnomer ''Darwin's finches,'' Marx & Bornmann, 2013; iMetrics, Leydesdorff, Bornmann, Marx, & Milojević, 2014; the Global Positioning System, Comins & Hussey, 2015). More recently, Comins & Hussey (2015a) expanded the original scope of this technique from analyzing a single RPYS output to allowing for the comparison and visualization of *multiple* outputs (referred to as multi-RPYS). Given the way standard RPYS is implemented, however, it remains unknown whether the publications driving these peaks result from *sustained* impact on the field over time as compared with, for instance, a short burst of citation activity indicative of a more transiently impactful paper. In this study, we raise the question of whether RPYS methodologies capture the sustained versus transient knowledge claims in a field's history.

To consider whether RPYS techniques identify papers of sustained impact, we begin by discussing the kinds of influences potentially revealed by RPYS analyses. Recent work by Baumgartner and Leydesdorff (2014) describes distinctions between the citation histories of articles that have a lasting impact on a research community--"citation classics" or, as the authors propose, "*sticky knowledge claims*"--contrasted with articles that show an initial burst of impact followed by a rapid decay within several years of publication-contributions at a research front or "*transient knowledge claims*". Standard RPYS techniques do not inherently contain a procedure to differentiate knowledge claims of substantial but temporary impact from those with a sustained influence over the field. In order to look for the representation of sticky knowledge claims in historical reconstructions generated by RPYS, one must operationalize how a sticky knowledge claim would appear in an RPYS analysis as compared with a transient knowledge claim. Baumgartner & Leydesdorff (2014) suggest that sticky knowledge claims "continue to be cited more than 10 years after publication" whereas transient knowledge claims decay within several years. The enduring relevance of sticky knowledge claims suggests these articles help define the identity of a research community and, as a result, the goal of using RPYS should be to identify such sticky claims.

Using both standard and multi-RPYS in tandem, we gauge whether sticky knowledge claims are depicted in a field's history when reconstructed from these procedures. While originally conceived to compare RPYS outputs across different topics, multi-RPYS provides an opportunity to similarly compare outputs within a single topic segmented over a variable, such as *time*. To test these ideas, we use a case study to compare the patterns of results seen from standard and multi-RPYS analyses of a single journal (*Scientometrics*). We performed multi-RPYS analyses by segmenting the journal's articles by time intervals over the past 37 years. This allows us to assess how consistently seminal works cited by articles from one era (e.g., articles published in the 1980s) are also referenced in other eras (e.g., articles published in the 1990s, 2000s, etc) and how these results correspond with standard RPYS outputs.

After evaluating the reliability of RPYS methods to identify those papers with an enduring relevance to a research community, we ask whether such information can be used to reconstruct relationships across research activities. To do so, we examine whether multi-RPYS analyses can be used to meaningfully cluster non-overlapping sets of publications by creating and comparing the intellectual histories derived from RPYS analyses of 40 journals from the field of information science and library science.

**2. Method**

Data was accessed and downloaded using the Thomson Reuters Web of Science (WoS) in September-November, 2015. First, we accessed the list of journals from the Web of Science Journal Citation Reports within the category of information science and library science. Journals were searched for within the Web of Science Core Collection. For each journal, results were refined by documents type to only include research articles. In total, we downloaded 37,649 records published in 40 of the information science and library science journals (the top 40 as ranked by Journal Impact Factor). Across these 40 journals, we identified 614,759 cited references to be used in our analyses. We note that within the information science and library science category, we aggregated results for *Journal of the Association for Information Science and Technology* and its name from 2001-2013, *Journal of the American Society for Information Science and Technology*.

For our initial case study analysis of the knowledge claims represented in RPYS, we organized articles from the library and information science journal *Scientometrics* at multiple time intervals: articles published between (1) 1978-1985, (2) 1986-1990, (3) 1991-1995, (4) 1996-2000, (5), 2001-2005, (6) 2006-2010 and (7) 2011-2015. First, we extracted cited references using a custom-written script. The script extracts publication years of references cited from the journal publication records. Next, reference publication years were combined by year to create a frequency distribution of references cited by a set of publications, where this distribution is sorted by the publication year of these cited references. Importantly, we note the potential of this routine to carry some error resulting from ambiguity of cited references and that approaches by other research groups are actively developing new techniques to improve cited reference disambiguation (Thor *et al.*, in preparation). For our case study we examine any cited references between 1900 and 2015 occurring within a *Scientometrics* article. For all subsequent journals analyzed in the second part of this study, we focus on references in articles published from 1900-1999.

For all analyses, we begin by taking the absolute deviation of the count of cited references in any given year (e.g., $x$) from the five-year median of the count of cited reference (e.g., $x$-2, $x$-1, $x$-2, $x$+1, $x$+2; Marx et al., 2014). To make the resulting vectors amenable to comparison, we rank transform them. Rank transformation was performed by taking the complete set of n data points from a given RPYS analysis and sorting them by magnitude.

These values are then substituted so that the largest takes the value n, the second largest takes n − 1, and so forth (Comins & Hussey, 2015a). Finally, to generate groupings based on shared historical origins, after journal RPYS analyses were rank-transformed, the data were submitted to Ward's minimum variance hierarchical clustering. In the second part of this paper, we compare furthermore observed RPYS results to a set of randomly generated RPYS results. To do so, we created a companion multi-RPYS analysis for the 40 journals analyzed in that section using *randomly-generated* values (i.e., ranks 1-100). We refer to this data set as the randomly-generated dataset.

## 3. Results

*3.1. Validating that constitutive citations are obtained in RPYS results: A case study of articles published in the journal Scientometrics*

We begin with a case study of the historical influences for articles published in the journal *Scientometrics* to determine if standard and mutli-RPYS can be used to verify the identification of sticky (long-lasting) as opposed to transient knowledge claims. Accordingly, we assess whether peak reference years in RPYS are (1) consistent over a time window of 10 or more years and (2) that specific papers repeatedly receive a large share of the references during that 10+ year timeframe.

Accordingly, we developed a case study allowing us to investigate the presence of sticky versus transient knowledge claims using RPYS. We performed RPYS analyses on a single representative journal from the field of quantitative science studies, *Scientometrics,* at multiple time intervals: articles published between (1) 1978-1985, (2) 1986-1990, (3) 1991-1995, (4) 1996-2000, (5) 2001-2005, (6) 2006-2010 and (7) 2011-2015. Thus, we label sticky knowledge claims in our multi-RPYS analysis if three or more time bins appeared highly referenced by a set of citing articles during ten or more years.

*Scientometrics* serves as a convenient dataset to assess the presence of sticky knowledge claims given previous research extensively examining historically important works emerging from a single RPYS analysis of all articles published in this journal (Leydesdorff et al., 2014). Importantly, while this prior research demonstrated the ability of the technique to identify seminal papers, it did not directly assess how *consistently influential* these seminal works were to the journal. For example, we note the following numbers of publications in our dataset over our 7 time intervals: (1) 297 articles published between 1978-1985, (2) 329 articles published between 1986-1990, (3) 406 articles published between 1991-1995, (4) 498 articles published between 1996-2000, (5) 511 articles published between 2001-2005, (6) 845 articles published between 2006-2010 and (7) 1,347 articles published between 2011-2015. In other words, nearly 52% of all *Scientometrics* articles have been published since 2006, which *could* overshadow the influence of references stemming from earlier journal publications.

To address this concern, we utilize a standard RPYS analysis of *Scientometrics* articles as well as a multi-RPYS analysis of those article segmented over time. The bottom of Figure 1 summarizes the results from multi-RPYS analyses of *Scientometrics* articles arranged by time of publication, while the top of Figure 1 contains the standard RPYS analysis of all articles. There are two initial observations to make of these results.

First, each analysis shows peak references for articles published within the span of that time interval. In other words, for the RPYS of articles published between 1978-1985, an increase is seen in references published in 1981 and 1982, for articles published between 1985-1990 an increase is seen in references published in 1985 and 1986, and so on. The increase in referencing results from articles citing more recent literature (Marx, 2011) and, more interestingly, their failure to persist across the time intervals analyzed suggests these represent transient knowledge claims within the analysis.

The second major observation is that, among the older references, one can identify a number of *bands* where reference years occur consistently across time intervals and we hypothesize such bands would be indicative of so-called sticky knowledge claims in multi-RPYS outputs (Baumgartner & Leydesdorff, 2014). We note that these bands correspond with the peaks observed in the standard RPYS analysis (see Figure 1, top) performed across the entire collection of articles published in *Scientometrics* (1978-2015) and are consistent with the notion that standard RPYS analyses can indeed capture the sticky knowledge claims in a field.

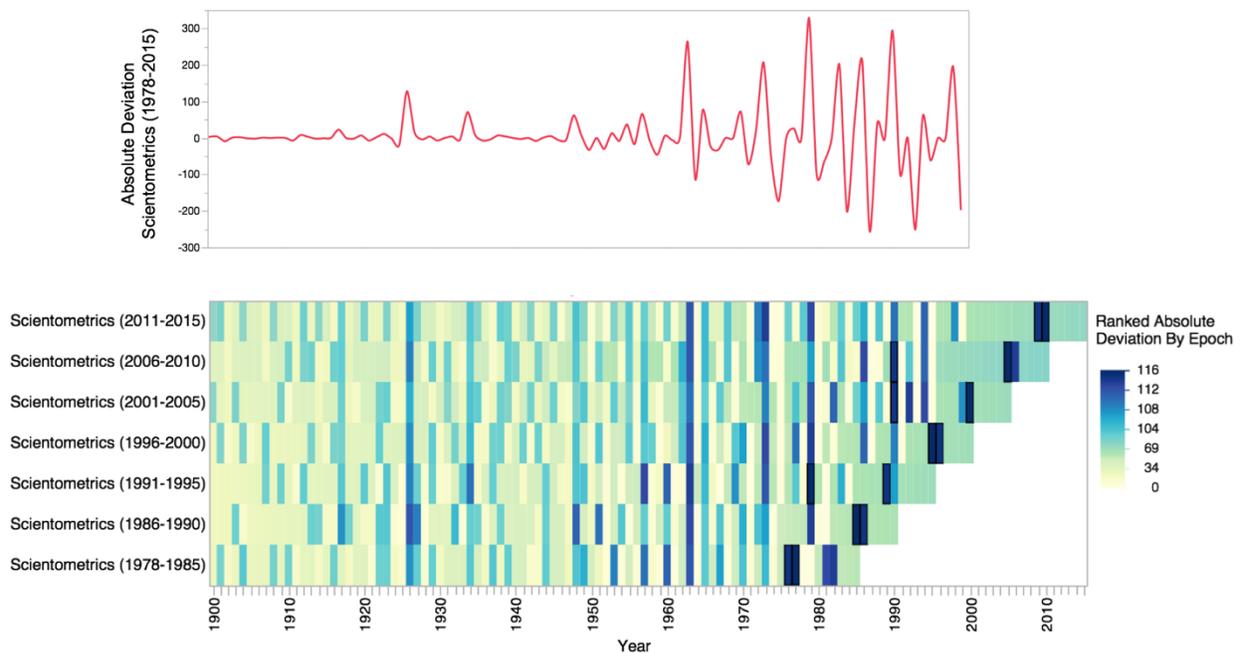

**Figure 1.** Standard (top) and multi-RPYS (bottom) analysis applied to articles published in the journal *Scientometrics.* The heatmap pertains to the multi-RPYS analysis performed over articles published in the journal *Scientometrics* during seven time intervals: (1) 1978-1985, (2) 1986-1990, (3) 1991-1995, (4) 1996-2000, (5), 2001-2005, (6) 2006-2010 and (7) 2011-2015. We observe several major bands, where historically important years are identified consistently across multiple time intervals corresponding to peaks in the standard RPYS analysis (see top). Further in-depth exploration of the major works underlying these bands is described in Table 1.

Let us explore the seven prominent bands, which last for at least a 10-year window, for consistency in terms of the underlying influential works corresponding to the years 1926, 1934, 1963, 1973, 1979, 1990 and 1994 (see Table 1). These years correspond with prominent bands observed in our heatmap:

- The peak in 1926 consistently refers to the publication of Lotka's law (Lotka, 1926) on the frequency of authors' publications within research communities.
- The second band occurs in 1934 and is driven by Bradford's "Sources of information on specific subjects" published in *Engineering* (Bradford, 1934).
- The next band in 1963 is predominantly driven by the classic *Little Science, Big Science* written by Derek de Solla Price, which discusses the exponential growth of science (de Solla Price, 1963).
- The band in 1973 has several major contributors, including Jonathan and Stephen Cole's *Social Stratification in Science* (Cole & Cole, 1973), Robert Merton's *The Sociology of Science: Theoretical and Empirical Investigations* (Merton, 1973) and Henry Small's article introducing co-citation networks (Small, 1973).
- The band in 1979 is driven from Eugene Garfield's book *Citation Indexing - Its Theory and Application in Science, Technology and Humanities* (Eugene Garfield, 1979) and Beaver and Rosen's work on scientific collaborations (de Beaver & Rosen, 1979a; de Beaver & Rosen, 1979b).
- Major works underlying the 1990 band are Egghe and Rousseau's *Introduction to Informetrics: Quantitative Methods for Library, Documentation and Information Science* (Egghe & Rousseau, 1990), Schubert and Braun's article "International Collaboration in the Sciences 1981-1985" (Schubert & Braun, 1990), as well as several articles on the relationship between patent metric and economic outcomes, including Trajtenberg's article entitled "A penny for your quotes: patent citations and the value of innovations" (Trajtenberg, 1990), Griliches' "Patent statistics as Economic Indicators: A Survey" (Griliches, 1990), Cohen and Levinthal's "Absorptive Capacity: a new perspective on learning and innovation" (Cohen & Levinthal, 1990).

- Finally, the 1994 band emerges from Wasserman and Faust's *Social Network Analysis: Methods and Applications* (Wasserman & Faust, 1994) and Gibbons et al.'s (1994) *The New Production of Knowledge*.

| Year Band | Analysis of *Scientometrics* articles published between 1978 and 2015 | | | | | | |
|---|---|---|---|---|---|---|---|
| | 1978-1985 | 1986-1990 | 1991-1995 | 1996-2000 | 2001-2005 | 2006-2010 | 2011-2015 |
| 1926 | Lotka: 85% | Lotka: 88% | Lotka: 100% | Lotka: 85% | Lotka: 89% | Lotka: 74% | Lotka: 98% |
| 1934 | Bradford: 60% | Bradford: 100% | Bradford: 81% | Bradford: 100% | Bradford: 73% | Bradford: 47% | Bradford: 66% |
| 1963 | Price: 44% | Price: 47% | Price: 52% | Price: 58% | Price: 46% | Price: 39% | Price: 35% |
| 1973 | Cole: 11% Merton: 6% Small: 4% | Cole: 10% Merton: 5% Small: 10% | Cole: 4% Merton: 13% Small: 12% | Cole: 6% Merton: 9% Small: 9% | Cole: 9% Merton: 12% Small: 20% | Cole: 15% Merton: 17% Small: 21% | Cole: 6% Merton: 12% Small: 27% |
| 1979 | | deB&R: 2% Garfield: 16% | deB&R: 8% Garfield: 9% | B&R: 11% Garfield: 24% | B&R: 9% Garfield: 16% | B&R: 8% Garfield: 24% | B&R: 9% Garfield: 20% |
| 1990 | | | C&L: 0% Egghe: 2% Griliches: <1% Schubert: 4% Trajtenberg: 0% | C&L: 0% Egghe: 6% Griliches: 1% Schubert: 4% Trajtenberg: <1% | C&L: 2% Egghe: 9% Griliches: 3% Schubert: 2% Trajtenberg: 0% | C&L: 3% Egghe: 5% Griliches: 6% Schubert: 3% Trajtenberg: 4% | C&L: 5% Egghe: 4% Griliches: 4% Schubert: 2% Trajtenberg: 6% |
| 1994 | | | | | Gibbons et al: 5% W&F: 2% | Gibbons et al: 4% W&F: 5% | Gibbons et al: 4% W&F: 9% |

**Table 1.** Description of the major works underlying bands for historically important reference years seen in Figure 1. Cells contain the percentage of all references attributable to a given work/author from *Scientometrics* articles published within the timeframe labeled in the corresponding column.

In summary, we show that, at least for the case of articles published in *Scientometrics*, RPYS analyses can contain representations of both transient and sticky knowledge claims. Transient claims consistently show up close to the most recent year of references collected and do not appear to linger in RPYS analyses after a period of 5-7 years. Sticky knowledge claims, on the other hand, are found reliably across multiple time intervals analyzed (subsequent to the reference's publication year). Thus, RPYS methods capture the enduring contributions of a research community that are vital to its identity and, as such, makes such methods attractive for assessing the similarity of multiple sets of research publications based on shared citation histories.

*3.2. Clustering documents with common intellectual histories: RPYS Analyses of Library & Information Science Journals*

We next evaluate the ability of RPYS methods for grouping sets of research articles together largely based on points of inflection in their shared intellectual histories. Using journals within the Web of Science Journal Category for information science and library science, we perform a multi-RPYS analysis by computing the standard time-normalized RPYS vector for each journal and then rank ordering those results. As a first step, we identified patterns present in our observed multi-RPYS results that were not found in a randomly-generated control dataset (see methods; Figure 2).

From this comparison, we note two key differences between the observed and randomly-generated data. First, results from the observed data show the potential for having at least some bands where the same years are identified as seminal across journals, whereas the randomly-generated data shows very weak evidence of the potential for heatmap bands of important years being consistent across journals. Second, the observed results reveal a major temporal bias for the years identified as seminal (which are primarily after 1960) whereas the randomly-generated data shows a seemingly uniform distribution of seminal years across the 100-year span of the analysis. Given the different behaviors in the observed versus randomly-generated data, we applied a hierarchical clustering procedure to our results to help organize which journals share similar patterns of important reference years.

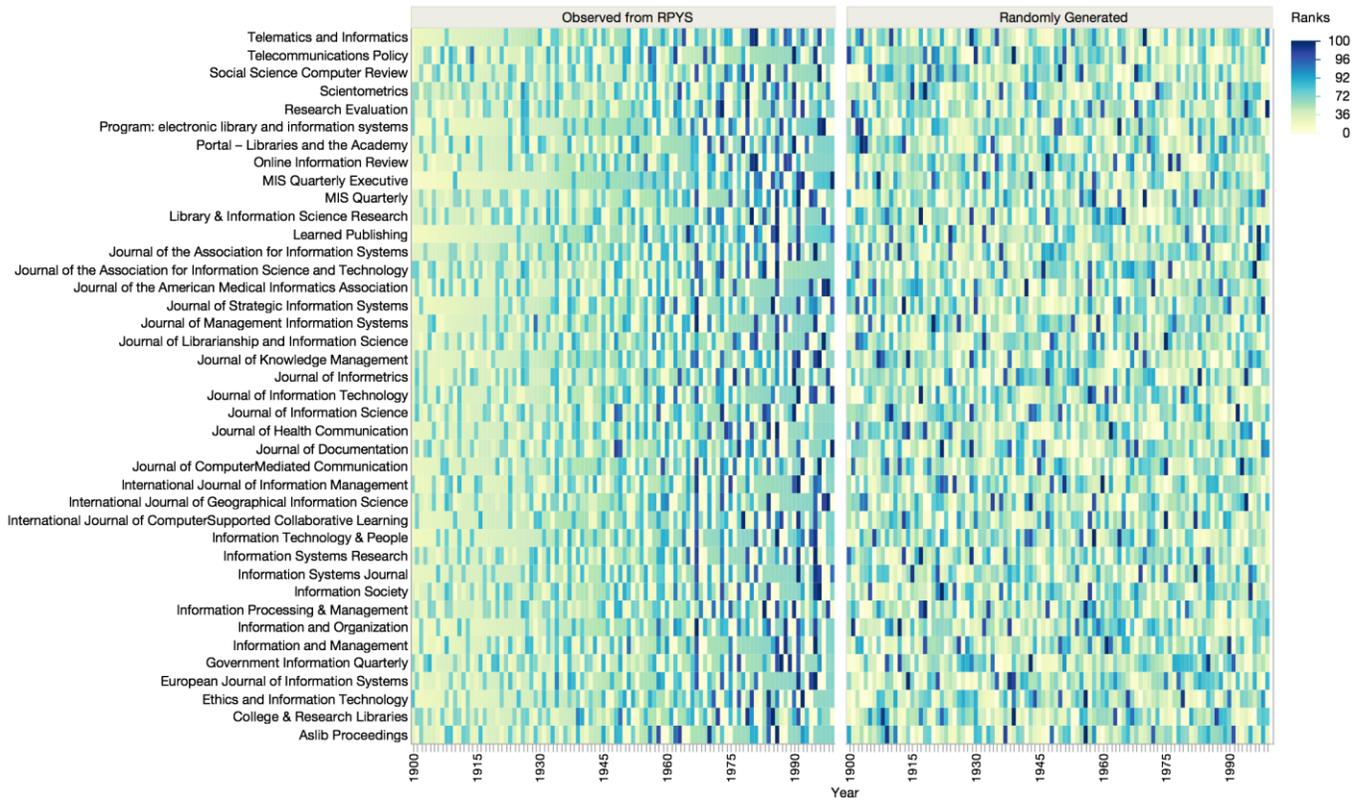

**Figure 2.** This figure shows the heatmap of multi-RPYS results observed for 40 journals published in the Journal Citation Reports category information science and library science (left) contrasted with randomly generated rank data for those same journals (right). Several notable differences are seen between the observed and randomly-generated data. Namely, for the figure on the left we find (1) the potential presence of bands where the same years are identified as seminal across journals and (2) a major bias in finding years identified as seminal after 1960. The randomly-generated data, on the other hand, shows very weak evidence of bands where important years are seen across journals and a uniform distribution of seminal years across the 100-year span of the analysis.

Submitting these RPYS results to hierarchical clustering analysis returns the taxonomy shown in Figure 3. The largest split observed separates information science and library science journals into two major groups. The first contains journals more focused on administrative and analytical aspects of information science. Within this category, there are two primary branches: librarianship and library systems (journals labeled "1" in Figure 3) and theoretical and applied information science (journals labeled "2" in Figure 3). Regarding the former, several of these journals allude to the role of the institution of libraries or librarians themselves (e.g., *College & Research Libraries*, *Portal - Libraries and the Academy*), while others focus on the discussion of library research and modern computing (e.g., *Program: Electronic Library and Information*

*Systems, Social Science Computer Review*) and the application of information science to certain specializations (e.g., *Ethics and Information Technology*, *Journal of Health Communication*, *International Journal of Geographical Information Science*). The latter collection includes journals covering theory and modeling of information science topics (e.g., *Information Processing & Management*, *Journal of Information Science*, *Journal of Documentation*, *JASIST*) as well as quantitative approaches to measuring and analyzing science and technology (e.g., *Scientometrics*, *Journal of Informetrics*, *Research Evaluation*).

The second major group contains journals with an emphasis on information and knowledge managements systems (Waltman, Yan, & van Eck, 2011), including topics focusing management systems (e.g., *European Journal of Information Systems*, *Information Systems Journal*, *Journal of Knowledge Management*) and the role of information science and technology in organizational environments (e.g., *Information Technology & People*, *International Journal of Computer-Supported Collaborative Learning*, *Journal of Computer-Mediated Communication*). We acknowledge here that, while this branch itself contains a division that separates four journals (journals labeled "4": *Government Information Quarterly*, *Journal of the Association for Information Systems*, *Telecommunications Policy*, *Journal of the American Medical Informatics Society*), they lack a thematic cohesion compared with groupings of other journals in our analyses.

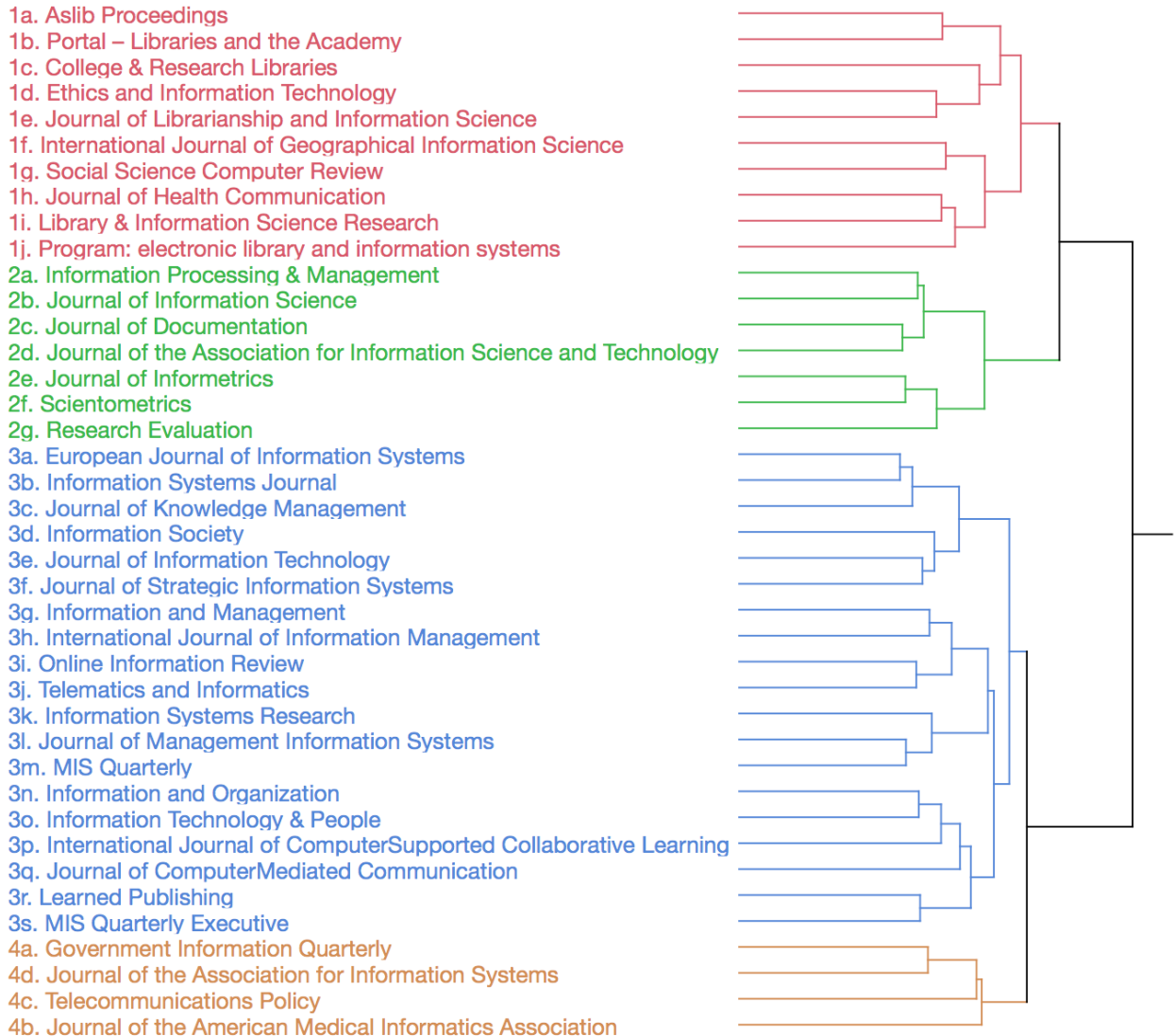

**Figure 3.** Hierarchical clustering of the multi-RPYS analysis performed on 40 journals published in the Journal Citation Reports category information science and library science. The two major branches of separation contain multiple internal clusters. Colors and numbers correspond to the largest levels of sub-clustering found within the two major branches.

Clustering these 40 journals from the Web of Science Journal Category for information science and library science using the sparse outputs of multi-RPYS analyses provided surprisingly meaningful, albeit imperfect, journal taxonomies. Figure 4 shows the resorted observed RPYS results from Figure 2 (left) according to the output of the hierarchical clustering procedure. This re-sorting (Figure 4) reveals clear patterns shared across the RPYS representations, an indicator of similarity across these journals (e.g., the presence of a band in 1926 across cluster 2). Importantly, we note that highly specialized journals that are somewhat unique within the set of journals we selected appear readily susceptible to miscategorization (e.g., *Journal of the American Medical Informatics Association*). Thus, it seems that multi-RPYS methods can identify the communality of historical roots, as can be seen for *Scientometrics* and *Journal of Informetrics*, journals with well-documented shared historical origins (Leydesdorff et al, 2014).

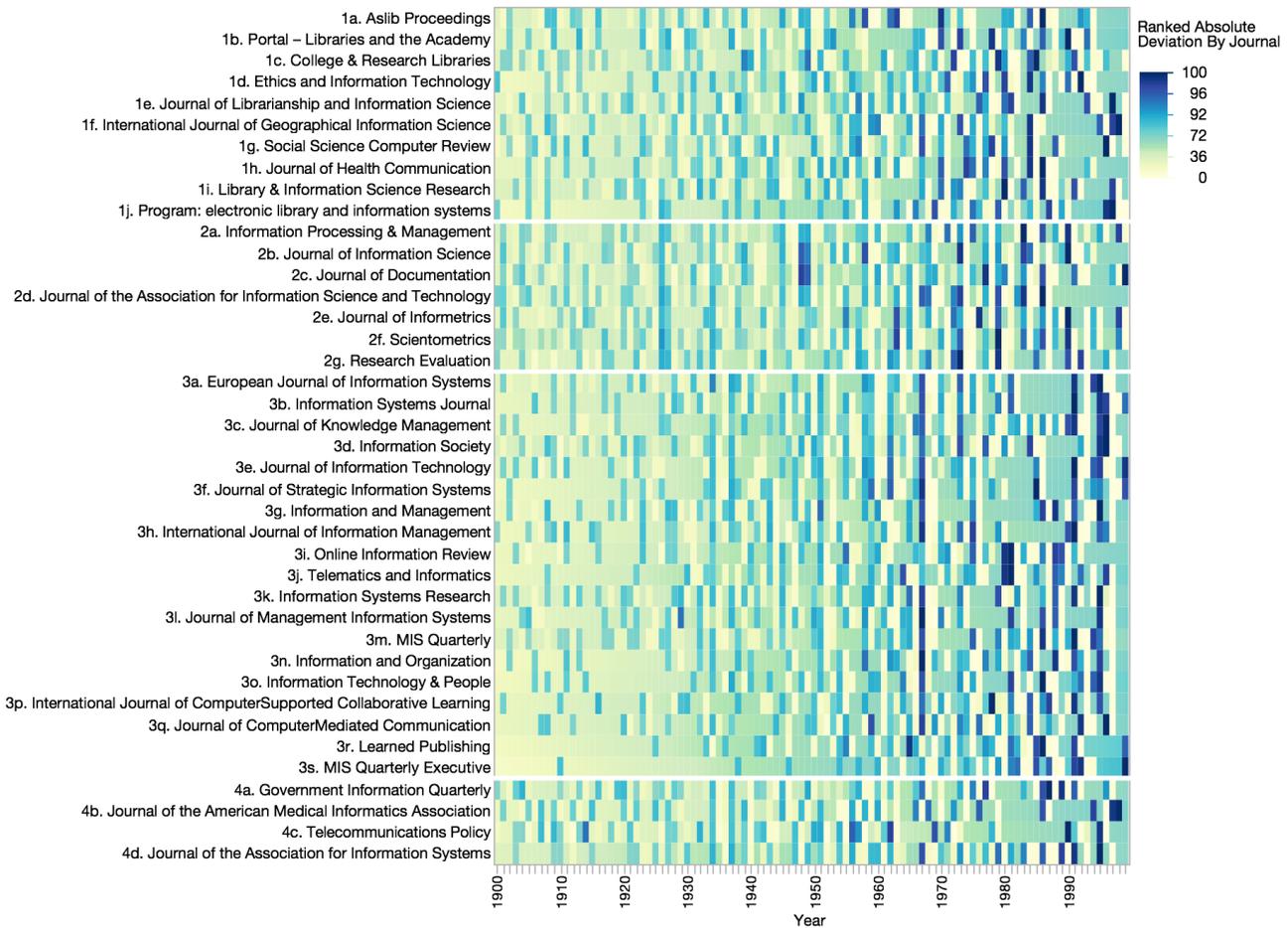

**Figure 4.** Here we use the results of our hierarchical clustering procedure to re-organize our heatmap of the multi-RPYS analysis (see Figure 2, left panel). The results yield clearly observable bands of impact across and within cluster and sub-cluster areas. For instance, a blue band can be seen across all journals in subcluster 2, suggesting that this year played an influential role in the scientific history of all seven of these journals.

## 4. Discussion

Developing data-driven techniques for classifying research fields and communities is a central goal in quantitative approaches to measuring and analyzing science and technology. Indeed, recent work strongly argues that field-normalization is an essential component of creating indicators capable of articulating the impact of prior scientific works and investigators (Leydesdorff & Bornmann, 2015; Ruiz-Castillo & Waltman, 2015; Waltman, van Eck, van Leeuwen, Visser, & van Raan, 2011). Here we assess whether the application of multi-RPYS methods, which are informed by the historical origins of the records being analyzed, can serve as proxies for helping to reconstruct relationships between sets of publications based on historical perspectives.

We began by using a case study to demonstrate that results emerging from RPYS analyses do indeed tend to be driven by enduringly important contributions to the field as opposed to contributions that reveal only temporary impact – a distinction that we referred to as "sticky" and "transient" knowledge claims (following Baumgartner & Leydesdorff, 2014). Surprisingly, despite only using sparse, rank-transformed representations of journal histories, we were able to create meaningful history-based journal taxonomies within the narrow context of journals from the Journal Citation Reports category information science and library science.

Importantly, more study and refinements are required to determine whether this procedure can be refined to eliminate errors, such was the case for the *Journal of the American Medical Informatics Association* in this study, or whether these errors are indicative of potentially strict limitations of this procedure. We recognize that applying this analysis as is to less well-constrained datasets (e.g., all scientific journals as opposed to journals within a particular pre-defined subject area) will likely yield more erroneous clustering due to the sparse nature of the data being analyzed and the increased likelihood of coincidence. Finally, we acknowledge the possibility that in our large 40 journal multi-RPYS, shared peak years across journals could have been driven by different publications. With those caveats in mind, future work should refine mapping relationships between citing sets based on communal intellectual histories using multi-RPYS so that they are applicable to much larger and more diverse datasets.

Within the context of developing algorithms to reconstruct a field's intellectual history, we are reminded that "the map is not the territory" (Korzybski, 1933). RPYS and multi-RPYS methods are merely one set of tools for generating algorithmic historiographies of scientific fields and such techniques will likely never provide a "perfect" representation of a field's origins. That being said, the results provided in this paper show two key points. First, multi-RPYS enables us to validate whether peaks occurring in standard RPYS methods are driven by short-term citation peaks at the research front that decay within ten years or by historically constitutive (long-term) citations. Second, one can cluster sets of citing documents on the basis of their shared intellectual histories using the outputs of RPYS analyses. In particular, the latter technique demonstrates an advantage of utilizing RPYS techniques in studying a field's history in addition to popular tools such as Eugene Garfield's *HistCite*, which enables researchers to identify citation classics (Garfield, Pudovkin, & Istomin, 2003; Garfield, 2004, 2009) but does not provide a methodology for comparing and clustering intellectual histories across different document sets. The procedures proposed by this paper show that RPYS methods not only retrieve the intellectual roots of document sets under study (cited references) but also explore relationships based on intellectually shared histories amongst the citing sets.